\documentclass[12pt,a4paper]{article}

\usepackage{amsmath}
\usepackage{amssymb}
\usepackage[utf8]{inputenc}
\usepackage[T1]{fontenc}
\usepackage[margin=10pt, font=small, labelfont=bf]{caption}
\usepackage[left=2.5cm, right=2.5cm, top=3cm, bottom=4cm]{geometry}
\usepackage{authblk}
\usepackage[english]{babel}
\usepackage{epsfig}
\usepackage{bm}

\title{Interaction-Range Effects and Universality in the BCS-BEC
Crossover of Spin-Orbit Coupled Fermi Gases}

\author[1]{D. Giambastiani}
\affil{Dipartimento di Fisica ``Enrico Fermi'', Universit\`a di Pisa and INFN, Largo B. Pontecorvo 3, I-56127 Pisa, Italy}
\author[2]{M. Barsanti}
\affil{Dipartimento di Ingegneria Civile e Industriale, Universit\`a di Pisa and INFN, Largo L. Lazzarino, I-56122 Pisa, Italy}
\author[1]{M. L. Chiofalo}

\date{}

\begin{document}

\maketitle

\begin{abstract}
We explore the evolution of a ultracold quantum gas of interacting fermions crossing from a Bardeen-Cooper-Schrieffer (BCS) superfluidity to a Bose-Einstein condensation (BEC) of molecular bosons in the presence of a tunable-range interaction among the fermions and of an artificial magnetic field, which can be used to simulate a pseudo-spin-orbit coupling (SOC) and to produce topological states. We find that the crossover is affected by a competition between the finite range of the interaction and the SOC and that the threshold $\lambda_B$ for the topological transition is affected by the interactions only in the small pair size, BEC-like, regime. Below $\lambda_B$, we find persistence of universal behavior in the critical temperature, chemical potential, and condensate fraction, provided that the pair correlation length is used as a driving parameter. Above threshold, universality is lost in the regime of large pair sizes. Here, the limiting ground state departs from a weakly-interacting BCS-like, so that a different description is required. Our results can be relevant in view of current experiments with cold atoms in optical cavities, where tunable-range effective atomic interactions can be engineered.
\end{abstract}

\section{Introduction}

The evolution of superfluidity in a quantum gas of interacting fermionic particles crossing over from a Bose-Einstein condensation (BEC) of composite bosons to a Bardeen-Cooper-Schrieffer (BCS)~\cite{Bardeen} superfluidity of Cooper pairs~\cite{Strinati_Report}, is still an open problem. The study of this smooth evolution between a BCS to BEC state of superfluidity, proposed first by Leggett~\cite{Leggett} and further explored by Nozi\`eres and Schmitt-Rink (NSR)~\cite{Nozieres}, has opened the way to a number of theoretical and numerical efforts aimed to describe the underlying physics via a universal parameter independent of the details of the interaction between fermions. After a celebrated log-log plot by Uemura {\it et al}.~\cite{Uemura}, in which a classification of different conventional and unconventional superconductors was proposed plotting the respective Fermi temperature vs.~the critical one, the BCS-BEC crossover has become a useful framework to better understand high-temperature superconductivity~\cite{Randeria}. In this context, Pistolesi and Strinati~\cite{Pistolesi} have introduced the pair-correlation length, representing the typical size of the fermion pairs, as the universal parameter to describe the crossover physics. In particular, they noticed that unconventional superconductors were characterized by shorter correlation lengths, though not entering the bosonic side of the crossover.

Quantum atomic gases~\cite{Cornell,Ketterle1,Debbie,Ketterle2} represent an ideal platform where the BCS-BEC crossover can be investigated, with the advantage of extreme precision and control typical of atomic physics. In fact, besides the tuning of temperature, quantum gases offer the possibility of tuning the dimensions from three down to zero, the interactions in strength and sign via the Fano-Feshbach mechanism~\cite{Fano} driven by magnetic or optical means, or else in range via the use of dipolar gases or Rydberg atoms, as well as accessing both internal, spin-like, and motional degrees of freedom~\cite{Tiesinga}. More recently, in Lev's group it has been demonstrated the possibility of tuning the range of effective interactions among atoms in optical cavities~\cite{Ben_Lev}. 

Current experimental advancements have opened the possibility of studying the BCS-BEC crossover of quantum gases, in the presence of an external interaction term formally equivalent to a (so-called artificial) magnetic field~\cite{Dalibard}, along with the possibility of realizing different kinds of couplings between atomic internal (e.g. spin or pseudo-spin variables) and external (e.g. momentum) degrees of freedom, via atom-light interactions induced by laser lights in bosonic~\cite{Spielman,Wu_onelaser} and fermionic \cite{Shanxi_SOC,Huang_twodimensional} systems. Quantum gases with artificial magnetic fields can be especially interesting to simulate the Fractional Quantum Hall effect~\cite{Spielman} whereas, with spin-orbit coupling, to simulate topological insulators and superconductors~\cite{Zhai}, and to boost applications for measurements of tiny forces, where the coupling between internal and motional degrees of freedom often represents a crucial tool. Therefore, quantum gases represent a very convenient experimental and theoretical laboratory to probe the BCS-BEC crossover idea against different microscopic models. 

Spin-orbit coupling (SOC) physics in the BCS-BEC crossover has been theoretically investigated for the first time by Vyasanakere and Shenoy {\it et al}.~\cite{Shenoy} in the presence of a contact interaction. They found that SOC induces the occurrence of a bound state, in both the BEC and BCS regimes, and leads to a change in the topology of the Fermi sphere. Among the different SOC types considered, the symmetric and oblate cases have been found the most interesting for the crossover physics with respect to the prolate setting. This topic has been the subject of further investigations of the critical temperature for superfluidity within the so-called pairing approximation originally introduced by Kadanoff and Martin~\cite{Kadanoff}, in which the non-condensed fermionic pairs are treated within a number-conserving scheme~\cite{Sopik}. The pairing approximation has been further developed and extensively used by Levin {\it et al.}~\cite{Levin} in connection with high-temperature superconductivity, to explain the occurrence of a pseudogap in the single-particle excitation spectrum in the form of a shift of spectral weight towards finite frequencies, and also to determine the spin and density response behavior of quantum gases with SOC in the presence of a contact interaction~\cite{Levin_SOC}.

In this paper, we address two open questions in the physics of interacting quantum gases with SOC in the symmetric case. First, motivated by current experiments with cold atoms in optical cavities~\cite{Ben_Lev}, we investigate the effects interactions tunable in strength and range. Second, we discuss the extent to which universality~\cite{Pistolesi} may persist in the crossover once SOC is introduced. We find that SOC and finite interaction range compete in determining the crossover physics: the former tends to create more tightly bound pairs with small size, while longer-range and stronger interactions favour the formation of larger-sized Cooper pairs. We also study how the threshold $\lambda_B$ of SOC strength~\cite{Shenoy}, above which the Fermi sphere undergoes a change of topology, is shifted while varying the interactions strength and range. As to the universal behavior, we find that this persists for small $\lambda$ values of the SOC coupling, provided that the correlation length is used as a driving parameter to embody both the strength and range of the atomic interactions~\cite{Pistolesi}. As expected instead, for larger $\lambda$ values well beyond the threshold for the topological change of the Fermi sphere, the fluid crosses over from a BEC-like regime towards a regime characterized by large-size pairs though significantly different from a weakly interacting BCS state. The persistence of universal behaviour in the crossover of fermions with SOC below the threshold for the topological transition, represents a relevant concept fostering a deeper understanding of the crossover paradigm and its microscopic implementations, which had not been evidenced so far.

The paper is organized as follows. After introducing the model and having defined the observables to be investigated, we present our results. We begin with the effect of the interaction strength and range on the behavior of the two main parameters of the theory, i.e. the correlation length and the SOC threshold strength $\lambda_B$. Then, the universal behavior of the observables is discussed. Finally, we summarize our concluding remarks and future perspectives.

\section{Model and self-consistent equations} \label{model}

We consider the Hamiltonian for a fluid of interacting fermionic atoms with an attractive potential $V$ and a symmetric Spin-Orbit-Coupling (S-SOC) with coupling strength $\lambda$. This can be cast in the compact form~\cite{Shenoy}:
\begin{equation}
\begin{split}
\mathcal{H}= & \sum_{\bm{k}} \psi_{\bm{k}}^{\dagger} \left[\left(\frac{\hbar^2 \bm{k}^2}{2m}-\mu \right)\mathbf{1} + \hbar^2 \lambda v_F \vec{\bm{\sigma}} \bm{k}\right] \psi_{\bm{k}} \\ - & \frac{1}{\mathcal{V}} \sum_{\bm{k},\bm{k}',\bm{q}} V_{\bm{k} \bm{k}'} c_{\bm{k}+\frac{\bm{q}}{2} \uparrow}^{\dagger} c_{-\bm{k}+\frac{\bm{q}}{2} \downarrow}^{\dagger} c_{-\bm{k}'+\frac{\bm{q}}{2} \downarrow} c_{\bm{k}'+\frac{\bm{q}}{2} \uparrow},
\label{HamSOC}
\end{split}
\end{equation}
where $\mathcal{V}$ is the volume, $m$ the fermion mass, $\mu$ the chemical potential, $\vec{\bm{\sigma}}$ is the bmor of Pauli matrices $\vec{\bm{\sigma}} \equiv (\sigma_x,\sigma_y,\sigma_z)$, $c$ and $c^{\dagger}$ are the destruction and creation fermionic operators, respectively, and $\psi_{\bm{k}} \equiv (c_{\bm{k}\uparrow}, c_{\bm{k}\downarrow})^T$ the Nambu-Gor'kov spinor. In our treatment, $V_{\bm{k} \bm{k}'}$ is a finite-range potential, modeled in a separable form as in Nozi\`eres and Schmitt-Rink~\cite{Nozieres}:
\begin{equation}
V_{\bm{k} \bm{k}'} = g \frac{4 \pi}{m k_0 k_F} w_{\bm{k}} w_{\bm{k}'},\qquad
w_{\bm{k}} = \sqrt{\frac{(k_0 k_F)^2}{(k_0 k_F)^2+ |\bm{k}|^2}}.
\label{separablepotential}
\end{equation}
Here and in the following, $k_F$, $v_F$ and $E_F$ are respectively the Fermi momentum, velocity and energy, and the system adimensional parameters of the theory are $g>0$ driving the atom interaction strength, $k_0^{-1}$ measuring the interaction range, and $\lambda$ the SOC coupling strength.  

In fact, the SOC term introduces the special motional-related direction $\bm{k}$, so that one may distinguish between parallel and antiparallel spins with respect to $\bm{k}$. The physics can thus be expressed in the so-called helicity basis:
\begin{equation}
\begin{cases}
& a_{\bm{k},+} = N_{\perp}^{+} (s_+ c_{\bm{k} \uparrow} + e^{-i\varphi_{\bm{k}}} c_{\bm{k} \downarrow}) \\
& a_{\bm{k},-} = N_{\perp}^{+} (e^{i\varphi_{\bm{k}}} c_{\bm{k} \uparrow} - s_+ c_{\bm{k} \downarrow})
\end{cases},
\label{helicitybasis}
\end{equation}
in terms of the size of the $k$-bmor $k_{\perp} \equiv \sqrt{k_x^2+k_y^2}$ perpendicular to $\hat{z}$, the relative amplitude $s_{\pm} \equiv (|\bm{k}| \pm k_z)/k_{\perp}$ of the parallel ($s_+$) and antiparallel ($s_-$) components with phase $\varphi_{\bm{k}} \equiv arg(k_x + ik_y)$, and overall normalization factor $N_{\perp}^{\pm} \equiv  {[2|\bm{k}|(|\bm{k}| \pm k_z)]^{-{1}/{2}}}/{k_{\perp}}$. This representation naturally leads to the description of the available fermionic states, in terms of two Fermi spheres of different helicity, i.e. different eigenvalues of the operator $\vec{\bm{\sigma}} \bm{k}$~\cite{Shenoy}. In particular, below a critical value $\lambda_T$ of the SOC parameter ($\lambda < \lambda_T$) the $-$ helicity sphere expands with increasing values of $\lambda$, while the $+$ helicity shrinks until it disappears at $\lambda = \lambda_T$. Above the threshold $\lambda_T$, a hole appears in the $-$ helicity Fermi sphere and, as a result, $-$ helicity fermions fill in the shell between two spheres~\cite{Shenoy}. 

In order to describe the physics at finite temperature including the non-condensed pairs, we resort to the mean-field conserving approximation introduced by Kadanoff and Martin~\cite{Kadanoff} and subsequently developed by Levin {\it et al.} under the name of pairing approximation~\cite{Levin}.
In our case with the separable potential $V_{\bm{k} \bm{k}'}$ and SOC in the helicity basis, the pairing approximation consists in solving the following self-consistent set of equations for the T-matrix $t(Q)$ embodying the response $\chi(Q)$, the single-particle self-energy $\Sigma_\alpha(K)$ and the interacting Green's function $G_\alpha(K)$ referring to helicities $\alpha=\pm$:   
\begin{flalign}
g= \frac{m k_0 k_F}{4\pi}[{1 + g\chi(Q)}]t(Q) &&
\label{T-matrix1}
\end{flalign}
\begin{flalign}
\chi(Q) = \frac{2 \pi k_B T}{m k_0 k_F \mathcal{V}} \, \sum_{K, \alpha} G_{\alpha}(K) G_{0 \, \alpha}(Q - K) w_{\bm{k} - \frac{\bm{q}}{2}}^2 &&
\label{T-matrix2}
\end{flalign}
\begin{flalign}
\Sigma_{\alpha}(K) & = \frac{k_B T}{\mathcal{V}} \sum_{Q} t(Q) G_{0 \, \alpha}(Q - K) w^2_{\bm{k}} \label{Sigma1} \\
& = G_{0 \, \alpha}^{-1}(K) - G_{\alpha}^{-1}(K)
\label{Sigma2} &&
\end{flalign}
\begin{flalign}
n = \frac{k_B T}{\mathcal{V}} \sum_{K, \alpha} G_{\alpha} (K), &&
\label{numberLevin}
\end{flalign}
Here, $n$ is the fermion density. We use the four-momentum $K\equiv (i\omega_n,\bm{k})$ and $Q\equiv (i\Omega_m,\bm{q})$, with $\omega_n=(2n+1)\pi k_BT/\hbar$ and $\Omega_m=2m\pi k_BT/\hbar$ the Matsubara frequencies for fermion and boson particles, respectively. The single-particle energy for each $\alpha$-helicity state is $\epsilon_{\bm{k}}^\alpha \equiv \hbar^2 \bm{k}^2/2m - \mu + \alpha \hbar^2 \lambda v_F |\bm{k}|$, entering the free fermion propagator $G_{0 \, \alpha} = (i\omega_n - \epsilon_{\bm{k}}^{\alpha})^{-1}$. 

Notice that the approximation expressed by eqs. (\ref{T-matrix1})-(\ref{numberLevin}) is not fully self-consistent, due to the appearance of the free $G_{0 \, \alpha}$ in (\ref{T-matrix2}) and (\ref{Sigma1}). In spite of this simplification, it is known~\cite{StefanucciLeeuwen,Sopik} that this approximation is capable to capture the main spectral properties of the system.  

The solution of the set (\ref{T-matrix1})-(\ref{numberLevin}) can be further simplified after resorting to the analysis developed by Levin {\it et al.}~\cite{Levin}, embodying all the relevant physics while being tested by means of full numerical calculations~\cite{Levin}. In essence, the T-matrix is decomposed into its singular component $t_{sf}$ related to the superfluid part, and a regular component $t_{pg}$ describing the pseudogap effects arising from the non-condensed pairs, thus 
\begin{equation}
t(Q)=t_{sf}(Q)+t_{pg}(Q). 
\label{ttotal}
\end{equation}
On the other hand, the
superfluid contribution can be approximated to be given by the BCS expression 
\begin{equation}
t_{sf}(Q) = -\frac{\Delta_{sf}^2}{k_B T}\delta(\bm{q})\delta_{i\omega_n, 0},
\label{tsf}
\end{equation}
with $\Delta_{sf}$ being the superfluid gap. Expression (\ref{tsf}) means that the fermion-fermion pairing contributes only through the condensate at zero-momentum. The superfluid self-energy can be expressed as:
\begin{equation}
\Sigma_{sf}(K) = \frac{k_B T}{\mathcal{V}} \sum_{Q, \alpha} t_{sf}(Q) G_{0 \, \alpha}(Q - K) w_{\bm{k}}^{2}, 
\end{equation}
so that, eventually, $\Sigma_{sf}$ takes the simple form
\begin{equation}
\Sigma_{sf}(K) = - \Delta^2_{sf} w_{\bm{k}}^2 \sum_{\alpha} G_{0 \, \alpha}(-K).
\label{sigma_sf}
\end{equation}
Inserting (\ref{ttotal}) and (\ref{tsf}) in (\ref{T-matrix1}), one obtains, when $Q \neq 0$
\begin{equation}
t_{pg}(Q)= \frac{4 \pi}{m k_0 k_F}\frac{1}{g^{-1} + \chi(Q)},
\label{tpg}
\end{equation}
and, by the same token, the Thouless criterion 
\begin{equation}
1 + g \chi(0) = 0,
\label{thouless}
\end{equation}
providing the gap equation. 

From (\ref{Sigma1}), the splitting of $t(Q)$ leads to a splitting of the self-energy into $\Sigma(K)=\Sigma_{sf}(K)+\Sigma_{pg}(K)$, with 
$\Sigma_{pg}(K) = (k_B T/V) \sum_{Q, \alpha} G_{0 \, \alpha}(-K+Q) w_{\bm{k}}^2 t_{pg}(Q)$. 
When considering the pseudogap effects in the vicinity of $T_c$, one may use the fact that the T-matrix, $t_{pg}$, is peaked around $Q=0$~\cite{Levin}, and obtain
\begin{equation}
\Sigma_{pg}(K) \approx \sum_{\alpha} G_{0 \, \alpha}(-K) w_{\bm{k}}^2 \frac{k_B T}{\mathcal{V}} \sum_{Q} t_{pg}(Q). 
\end{equation}

As a result, in analogy to the superfluid gap $\Delta_{sf}$, one may thus define the pseudogap parameter $\Delta_{pg}$, which in the vicinity of $T_c$ takes the form
\begin{equation}
\Delta_{pg}^2 = - \frac{k_B T}{\mathcal{V}} \sum_{Q \neq 0} t_{pg}(Q),
\end{equation}
in terms of the pseudogap T-matrix. The single-particle excitation energy depending on the helicity $\alpha$ results to be $E_{\bm{k}}^\alpha = \sqrt{(\epsilon_{\bm{k}}^{\alpha})^2 + \Delta_{\bm{k}}^2}$, with $\Delta_{\bm{k}} \equiv \Delta w_{\bm{k}}$ and $\Delta^2 = \Delta_{sf}^2 + \Delta_{pg}^2$ decomposed into its superfluid and pseudogap parts. The full fermion propagator can eventually be written as:
\begin{equation}
G_{\alpha}(K) = \frac{i \hbar \omega_n + \epsilon_{\bm{k}}^{\alpha}}{(i \hbar \omega_n)^2 - (E_{\bm{k}}^{\alpha})^2}.
\end{equation}

We are now in a position to express the gap and number equations including the non-condensed pairs. 
From the Thouless criterion in eq.~(\ref{thouless}) one obtains:
\begin{equation}
1 = g \frac{2 \pi}{m k_0 k_F \, \mathcal{V}} \sum_{\bm{k}, \alpha} \frac{1-2f(E_{\bm{k}}^{\alpha})}{2 E_{\bm{k}}^{\alpha}} w_{\bm{k}}^2,
\label{gap}
\end{equation}
with $f(x)$ the Fermi distribution function. On the other hand, from eq.~(\ref{numberLevin}) one has:
\begin{equation}
n = \frac{1}{\mathcal{V}} \sum_{\bm{k}, \alpha} \left[ \frac{1}{2} \left( 1-\frac{\epsilon_{\bm{k}}^{\alpha}}{E_{\bm{k}}^{\alpha}} \right) + \frac{\epsilon_{\bm{k}}^{\alpha}}{E_{\bm{k}}^{\alpha}} f(E_{\bm{k}}^{\alpha}) \right].
\label{number}
\end{equation}
At the critical temperature $T_c$, one has $\Delta(T=T_c) = \Delta_{pg}$. After expanding the pseudogap T-matrix up to first order in $i\Omega_m$ and second order in $\bm{q}$, one can derive an analytical expression for the pseudogap parameter~\cite{He}. In fact, the expansion coefficients can be expressed as:
\begin{equation}
\begin{split}
Z & =\frac{\partial t^{-1}(Q)}{\partial (i \hbar \Omega_m)} \Bigg|_{Q = 0}
\\
\frac{Z}{2 m_b} & =\frac{\partial^2 t^{-1}(Q)}{\partial \bm{q}^2} \Bigg|_{Q = 0},
\end{split}
\label{mbZ}
\end{equation}
and measure the renormalized bandwidth $Z$ and effective mass $m_b$ of the non-condensed resonant pairs. In terms of these quantities, one finally has the expression for the pseudogap $\Delta_{pg}$~\cite{He}:
\begin{equation}
\Delta_{pg}^2 = \frac{1}{Z} \left( \frac{T m_b}{2 \pi} \right)^{\frac{3}{2}} \zeta \left(\frac{3}{2} \right), 
\label{pseudogap_eq}
\end{equation}
with $\zeta$ the Riemann function. 

In the zero temperature limit with $\Delta_{pg}=0$ instead, eqs. (\ref{gap}) and (\ref{number}) can be cast in the form:
\begin{subequations}
\begin{equation}
1 = g \frac{2 \pi}{m k_0 k_F \mathcal{V}} \sum_{\bm{k}, \alpha} \frac{w_{\bm{k}}^2}{2 E_{\bm{k}}^{\alpha}},
\label{zero_gap}
\end{equation}
\begin{equation}
n = \frac{1}{2 \mathcal{V}} \sum_{\bm{k}, \alpha} \left( 1-\frac{\epsilon_{\bm{k}}^{\alpha}}{E_{\bm{k}}^{\alpha}} \right).
\label{zero_number}
\end{equation}
\end{subequations}
Once the gap and chemical potential are calculated, the condensate fraction $N_c$ can be worked out from~\cite{condensate_fraction}:
\begin{equation}
N_c = \frac{3 \pi^2 \Delta^2}{4 \mathcal{V}} \sum_{\bm{k}, \alpha} \left( \frac{w_{\bm{k}}}{E_{\bm{k}}^{\alpha}} \right)^2.
\label{zero_condensate}
\end{equation}

In conclusion, the solution of eqs. (\ref{zero_gap}), (\ref{zero_number}) and (\ref{zero_condensate}) provides the chemical potential $\mu(T=0)$ and superfluid gap $\Delta(T=0)$, along with the condensate fraction $N_c$ at $T=0$. The self-consistent solution of eqs. (\ref{gap}), (\ref{number}) and (\ref{pseudogap_eq}), provide the critical temperature $T_c$ along with the pseudogap $\Delta_{pg}(T_c)$ and chemical potential $\mu(T_c)$.

\section{Correlation length}\label{observables}
Before discussing the numerical results in the crossover, we need a quantitative measure of BCS-like and BEC-like behaviors at $T=0$. This is provided by the correlation length $\xi$, in essence the average width of the pair-correlation function. In fact, we define the BCS limit of the crossover as that for which $k_F \xi \gg 1$ corresponding to large-size pairs, while the BEC limit as that with $k_F \xi \ll 1$, that is small-size pairs.

In the following, we calculate $\xi$ at $T=0$ in the crossover. To this aim, we first determine the BCS-like ground state in the very convenient helicity basis (\ref{helicitybasis}). In terms of the indexes $\sigma$ and $\sigma'=\uparrow,\downarrow$, this is:
\begin{equation}
| \Theta \rangle = \prod_{\bm{k}} \frac{1 + \sum_{\sigma, \sigma'} \phi_{\sigma \sigma'}(\bm{k})c_{\bm{k}\sigma}^{\dagger}c_{-\bm{k}\sigma'}^{\dagger}}{\sqrt{1+\sum_{\sigma \sigma'}|\phi_{\sigma\sigma'}(\bm{k})|^2}} |0\rangle,
\label{groundstate}
\end{equation}
where
\begin{equation*}
\begin{split}
& \phi_{\uparrow \downarrow} (\bm{k}) = - \phi_{\downarrow \uparrow} (\bm{k}) = - \frac{1}{2} \Big(\phi_{\bm{k}}^{+} s_+ + \phi_{\bm{k}}^{-} s_- \Big) \frac{k_{\perp}}{|\bm{k}|},
\\ &\phi_{\uparrow \uparrow}(\bm{k}) = -\phi_{\downarrow \downarrow}^{\ast} (\bm{k}) = \frac{1}{2} \Big( \phi_{\bm{k}}^{+} - \phi_{\bm{k}}^{-} \Big) \frac{k_{\perp}}{|\bm{k}|} e^{-i \varphi_{\bm{k}}}.
\end{split}
\end{equation*}
Here, $\varphi_{\bm{k}} \equiv arg(k_x + ik_y)$, as defined before, we have defined $\phi_{\bm{k}}^{\alpha} = v_{\bm{k}}^{\alpha}/u_{\bm{k}}^{\alpha}$ and
\begin{equation*}
(u_{\bm{k}}^{\alpha})^2 = \frac{1}{2} \left( 1 + \frac{\epsilon_{\bm{k}}^{\alpha}}{E_{\bm{k}}^{\alpha}} \right) \,\,\, \text{and} \,\,\, (v_{\bm{k}}^{\alpha})^2 = \frac{1}{2} \left( 1 - \frac{\epsilon_{\bm{k}}^{\alpha}}{E_{\bm{k}}^{\alpha}} \right)
\end{equation*}
playing the role of the BCS $u_{\bm k}$ and $v_{\bm{k}}$ in the helicity basis labeled by $\alpha$. 

Now, we may introduce the pair distribution function:
\begin{equation*}
g_{\sigma \sigma'}(\bm{r} - \bm{r}') = \frac{1}{n^2} \langle \Theta | [ \hat{\rho}_{\sigma}(\bm{r}) \hat{\rho}_{\sigma'}(\bm{r'}) - \delta_{\sigma \sigma'} \delta(\bm{r}-\bm{r}') \hat{\rho}_{\sigma}(\bm{r}) ] | \Theta \rangle,
\end{equation*}
normalized to the squared average density $n$ of the system, with $\hat{\rho}_{\sigma}(\bm{r})= c_{\sigma}^{\dagger}(\bm{r}) c_{\sigma}(\bm{r})$ being the density operator for particles with spin $s$ and position $\bm{r}$. At variance with the conventional BCS case without SOC, here the pair-correlation function is in general characterized by non-zero amplitudes in both the singlet and triplet channels. Thus, we define 
$g(\bm{r}) \equiv \sum_{\sigma \sigma'}g_{\sigma \sigma'}(\bm{r})$ and the corresponding (squared) pair correlation length as $\xi^2 = [\int \bm{r}^2 g(\bm{r}) d\bm{r}]/[\int g(\bm{r}) d\bm{r}]$.
After Fourier transformation and neglecting - as it is customary - Hartree and Fock contributions~\cite{Strinati_heliash}, we obtain the expression:
\begin{equation}
\xi^2 = \frac{1}{\mathcal{N}}\int d\bm{k} \sum_{\sigma, \sigma'} \left| \nabla_{\bm{k}} A_{\bm{k}}^{\sigma, \sigma'} \right|^2 ,
\label{xi2}
\end{equation}
with $\mathcal{N}\equiv{\int d\bm{k}  \sum_{\sigma, \sigma'} \left| A_{\bm{k}}^{\sigma, \sigma'} \right|^2 }$. Eq.~(\ref{xi2}) generalizes the BCS expression for $\xi$~\cite{Pistolesi} to the inclusion of SOC, with: $A_{\bm{k}}^{\sigma, \sigma'} \equiv {\phi_{\sigma, \sigma'}(\bm{k})}[1 + \sum_{\sigma \sigma'}|\phi_{\sigma \sigma'}(\bm{k})|^2]^{-1}$.
In the following, we discuss the behavior of the $T=0$ (from eqs. (\ref{zero_gap}), (\ref{zero_number}) and (\ref{zero_condensate})) and $T=T_c$ (from eqs (\ref{gap}), (\ref{number}) and (\ref{pseudogap_eq})) quantities in the crossover spanned by $k_F\xi$ (as calculated from (\ref{xi2})), while varying the strength $g$ and range $k_0^{-1}$ of the pairing interaction, and the SOC strength $\lambda$. To this aim, we first analyze the behavior of the pair correlation length and of the threshold of $\lambda$ above which the change of topology occurs in the Fermi sphere. 

\section{Correlation length and threshold SOC strength: competition between SOC and finite-range effects}\label{resultsxilambdaT}
\begin{figure}
\centering
\includegraphics[scale=0.5]{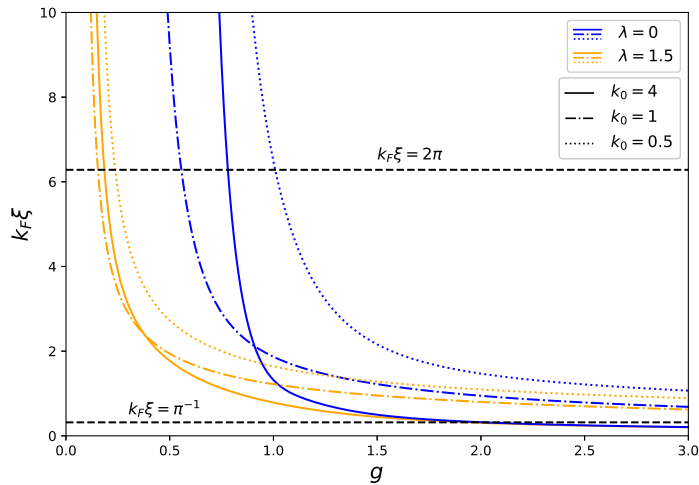}
\caption{Pair correlation length $k_F\xi$ vs.~the coupling $g$, without (blue lines) and with (yellow lines) SOC, for different interaction ranges as in the legend: $k_0 = 0.5$ (dotted line), $k_0 = 1$ (dotted-dashed line) and $k_0 = 4$ (solid line). The horizontal dashed lines $k_F\xi=2\pi$ and $k_F\xi=\pi^{-1}$ correspond to the thresholds for deep BCS-like ($k_F\xi>2\pi$) or deep BEC ($k_F\xi<\pi^{-1}$) behavior in the crossover without SOC~\cite{Pistolesi}.}
\label{length}
\end{figure}

Fig.~\ref{length} displays the calculated correlation length $k_F \xi$ as a function of $g$ for different values of the SOC strength $\lambda$ (different colours as in the legend) and $k_0$ (different line types). We notice that, at fixed $g$, the introduction of SOC (yellow lines) favors a BEC-like regime with smaller pair sizes with respect to the absence of SOC (blue lines). Let us comment on the effect of finite interaction range. In the absence of SOC, longer interaction ranges (dotted-dashed lines) on the scale of $k_F^{-1}$, are seen to favor a BCS-like regime with larger pair sizes while $g > 1$. For weaker coupling values $g < 1$ than the threshold $g \approx 1$, the behavior is inverted and shorter ranges correspond to more BCS-like character. Once the SOC is switched on, the threshold of $g$ where the inverted behavior occurs, is lowered towards weaker coupling strengths, confirming the role of SOC in extending the BEC region of the crossover.      
From Fig.~\ref{length}, we can argue that in the presence of stronger attractive coupling $g$, the formation of larger size fermion pairs is favored by the action of longer rather than shorter interaction ranges, and viceversa for weaker coupling values of $g$. When SOC is switched on, the behavior at weaker coupling strengths is evidently flattened towards a BEC-like regime.
As expected from eq.~(\ref{groundstate}), SOC is seen to favor the emergence of triplet pairing with amplitudes $\phi_{\uparrow \uparrow}$ and $\phi_{\downarrow \downarrow}$, confirming the predictions from Ref.~\cite{Shenoy}. In addition, we find that the triplet amplitude is enhanced in the limit of weak coupling strengths in accord with~\cite{Shenoy} and, consistently with the effect of $k_0$ emerged so far, also for longer ranges of the interactions.
The second parameter of the theory is the SOC strength.
One question is therefore to which extent the threshold for
the change of topology in the Fermi sphere is affected by
the interactions. As detailed in Fig. 2 of the Supplemental material~\cite{SM}, we find that $\lambda_B$ is unchanged while in the deep
BCS regime with $k_F\xi\gtrsim 2 \pi$, independently of the interaction range, while the threshold $\lambda_B$ tends to be slightly lowered in the intermediate crossover region before entering the deep BEC limit. Longer-range interactions tend to contrast the effect of SOC in the stronger coupling regime with smaller $k_F\xi$, so that for the case with $k_0=0.5$, $\lambda_B$ is enhanced with respect to the ideal system in a small region just below $k_F\xi = 2 \pi$.

In the following, the behavior of $\xi$ and $\lambda_B$ is used to frame the parameters region for which universal behavior occurs in the crossover. 

\section{Universality} \label{results}
We now turn to discuss the behavior of the observables at $T=0$ and $T=T_c$. 
We first notice that the observed competition behavior is found also while analyzing the results for the superfluid gap $\Delta$ and condensate fraction $N_c$ at $T=0$ as functions of $\lambda$ for different values of $k_0$ in the weak-coupling regime with $g<1$. Larger values of $\Delta$ and $N_c$ expected for BEC-like behavior, characterize the system with larger SOC and shorter-range interactions at fixed $g$, while for weak SOC strengths $N_c$ and $\Delta$ have smaller values.

\begin{figure}
\centering
\includegraphics[scale=0.6]{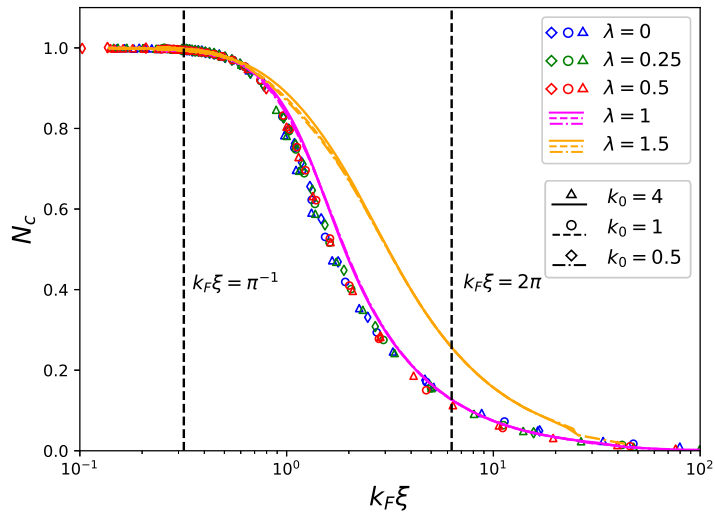}
\caption{Universal behaviour of the condensate fraction $N_c$ at $T=0$ vs.~the pair correlation length $k_F\xi$, varied after changing $g$ with $k_0$ and $\lambda$, as in the legend. Given the value of $\lambda_B$, in terms of the correlation length in Fig. 2 of the Supplemental material~\cite{SM}. For $\lambda < \lambda_B$, the values of $k_0$ are represented by different symbols: $k_0 = 0.5$ (squares), $k_0 = 1$ (circles) and $k_0 = 4$ (triangles). For $\lambda > \lambda_B$, the values of $k_0$ are represented by different line types: $k_0 = 0.5$ (dotted line), $k_0 = 1$ (dashed line) and $k_0 = 4$ (solid line). The coupling $g$ is in the range $0.05 < g < 50$. The vertical dashed lines at $k_F\xi=\pi^{-1}$ and $k_F\xi=2\pi$ signal the deep BEC and BCS thresholds, respectively. The value of $\lambda_B$ is obtained from Fig. 2 in the Supplemental material~\cite{SM}}
\label{Universal_condensate}
\end{figure}

In fact, the dependence of the condensate fraction, chemical potential, and critical temperature, on the different parameters appears to be greatly simpliflied once it is analyzed in terms of the correlation length $k_F\xi$. The degree of simplification is striking when compared with the behavior of the same quantities in terms of $g$, $k_0$, and $\lambda$ separately, as displayed in the Supplemental material~\cite{SM}. This is displayed in Fig.~\ref{Universal_condensate}, where all the data sets corresponding to different values of $g$, $k_0$ and $\lambda$, are collected and shown vs.~the corresponding $k_F\xi$. A striking universal behavior is observed for weak SOC strengths $\lambda\leq \lambda_B$, crossing over from the BEC regime with small $k_F\xi<\pi^{-1}$ to the BCS regime with large $k_F\xi>2\pi$. This universal behavior is however lost - not surprisingly - while $\lambda>\lambda_B$ and the Fermi sphere is changing its topology: here, the formation of the BEC is seen to be enhanced at a faster rate than for smaller SOC strengths, as expected. 

As to the chemical potential, the universal behavior is better seen after definining the effective quantity $\tilde{\mu} = \mu + m (\lambda v_F)^2/2$, i.e. the chemical potential measured from the bottom of the energy of the helicity state with lower energy. 
\begin{figure}
    \centering
    \includegraphics[scale=0.5]{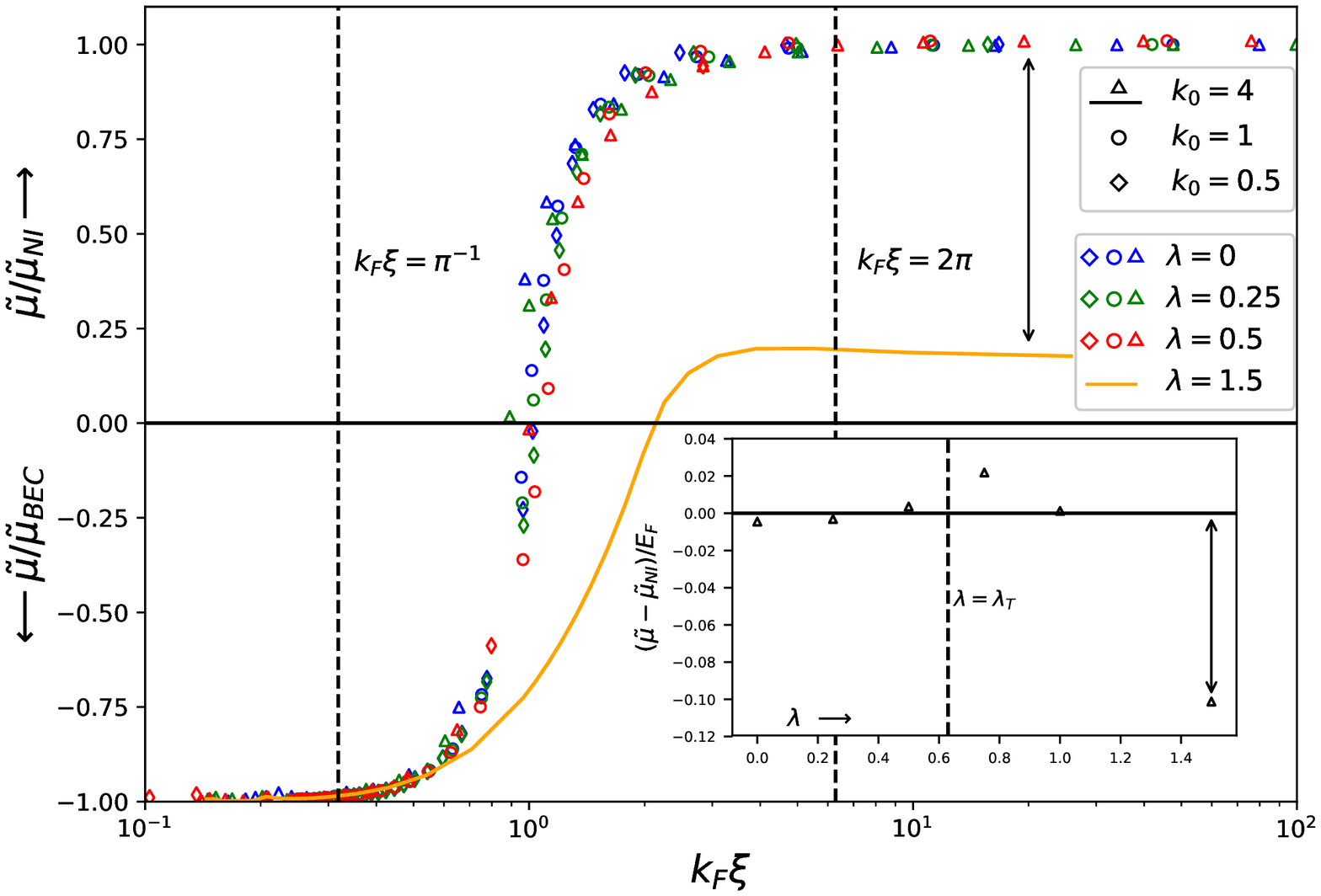}
    \caption{Universal behaviour of the effective chemical potential $\tilde{\mu}$ (see text) vs.~the pair correlation length $k_F\xi$, varied after changing $g$ with $k_0$ and $\lambda$, as in the legend.} For $\tilde{\mu} <0$ on the BEC side, $\tilde{\mu}$ is normalized to the limiting value $\tilde{\mu}_{BEC}$, that is half the pair binding energy. For $\tilde{\mu}>0$, $\tilde{\mu}$ is normalized to its non-interacting limiting value $\tilde{\mu}_{NI}$ (see text). Symbols are as in Figure \ref{Universal_condensate}. The coupling $g$ is in the range $0.05 < g < 50$. For $\lambda>\lambda_B$ the universal behavior is lost: here, the case with $\lambda=1.5$ and $k_0 = 4$ is shown for comparison. The vertical dashed lines at $k_F\xi=\pi^{-1}$ and $k_F\xi=2\pi$ signal the deep BEC (small pair sizes) and BCS (large pair sizes) thresholds, respectively. Inset: Zoom into the region  with $10.6<k_F \xi<12.1 $. Difference between $\tilde{\mu}$ and its corresponding non-interacting value $\tilde{\mu}_{NI}$, as a function of $\lambda$: for large $\lambda$, pair correlation lengths $k_F\xi > 2 \pi$ do not correspond to a weakly interacting BCS regime, independently of the interaction range (here, $k_0 = 4$).
    \label{fig:my_label}
\end{figure}
The effective chemical potential $\tilde{\mu}$ is displayed in Fig.~\ref{fig:my_label} vs.~$k_F\xi$, varied after changing $g$, $k_0$ and $\lambda$, as in the legend. When $\tilde{\mu} <0$, $\tilde{\mu}$ is normalized to the limiting value $\tilde{\mu}_{BEC}$, that is half the pair binding energy. This is obtained after numerical solution of the expression $1 = gk_0 [c^2 (1+ k_0) + \lambda^2]/\{(16c)[(c+k_0)^2 + \lambda^2]\}$, where $c^2 = - \tilde{\mu}/E_F$.
When $\tilde{\mu}>0$, $\tilde{\mu}$ is normalized to the non-interacting limiting value $\tilde{\mu}_{NI}$ given by $\tilde{\mu}_{NI}/{E_F}=2^{{1}/{3}}  \lambda^2/F(\lambda)+(F(\lambda)/2)^{1/3}-2\lambda^2$, with $F(\lambda)\equiv 1 + 2 \lambda^6 + \sqrt{1 + 4 \lambda^6}$~\cite{Li_He}.
A striking universal behavior persists while $\lambda$ is below the threshold $\lambda_B$ for the topological change of the Fermi sphere, extending the results in ~\cite{Pistolesi} to the inclusion of weak SOC interactions. 
Once again however, the data for $\lambda > \lambda_B$ fail to fall in the universal curve, essentially because $\tilde{\mu}$ does not stick to the BCS-like non-interacting limit. As it often occurs, the failure of universal behavior is far more evident in the chemical potential than in the condensate fraction. This dramatic change in behavior is zoomed in the inset of Fig.~\ref{fig:my_label}, where the difference between the effective chemical potential and its non-interacting counterpart is displayed for values of $k_F\xi$ well within the deep weakly-interacting BCS regime as a function of $\lambda$. In fact, $\tilde{\mu}$ is seen to significantly depart from $\tilde{\mu}_{NI}$. We thus observe that, though the pairs are on average characterized to be of large size, the weakly-interacting BCS state is not any-longer occurring well above the topological change. As a result, we infer that we do not have a crossover from a BCS to a BEC state but more properly from an effective BCS state of interacting bosonic molecules made of fermions with same helicity to an usual BEC state, confirming the results obtained by Yamaguchi {\it et al.}~\cite{Yamaguchi}.

\begin{figure}
\centering
\includegraphics[scale=0.6]{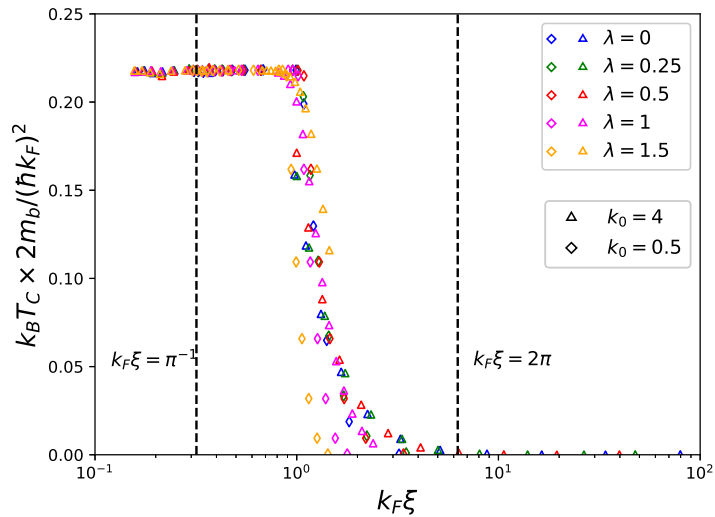}
\caption{Universal behaviour of the critical temperature $T_c$ times the effective mass $m_b$ of the non-condensed resonant pairs as a function of the pair correlation length $k_F\xi$ in log scale, for different values of $\lambda$ as in the legend. Symbols refer to different values of $k_0 = 0.5$ (squares) and $k_0 = 4$ (triangles). The coupling $g$ is in the range $1.25 < g < 30$. The vertical dashed lines at $k_F\xi=\pi^{-1}$ and $k_F\xi=2\pi$ signal the deep BEC and BCS thresholds, respectively. }
\label{Tc}
\end{figure}

Finally, we show in Fig.~\ref{Tc} the critical temperature $T_cm_b$ as a function of $k_F\xi$. Again, a striking universal behavior persists as long as $\lambda<\lambda_B$. Notice that the collapse of all the data with different values of $g$, $k_0$, and $\lambda$ occurs only after embodying the effective-mass correction arising for the non-condensed resonant pairs in eq.~(\ref{mbZ}). The data sets with $\lambda>\lambda_B$ are seen to depart from the universal curve, to an extent similar to the behavior of $N_c$ rather than that of $\tilde{\mu}$.

\section{Conclusions}\label{conclusions}

We have studied the effects of light-induced spin-orbit coupling (SOC) on the crossover from a Bardeen-Cooper-Schrieffer (BCS) to a Bose-Einstein Condensation (BEC) type of superfluidity in an interacting quantum gas of fermionic atoms in two different spin states, including at mean-field level the effects of the non-condensed pairs via the pairing approximation developed by ~\cite{Kadanoff,Levin}. We have focused on the so far unexplored effects obtained by tuning the range of the inter-atomic interactions and the degree of universal behaviour in the crossover. 

In particular, we find that in strong coupling conditions, a competition sets in between the SOC interaction, which moves the system towards the BEC side, and longer interaction ranges, which move the system towards the BCS side. At weak coupling, stronger SOC and longer range interactions cooperate to move the system towards the BEC side. In addition, we observe that universal behavior persists in the condensate fraction $N_c$ and effective chemical potential $\tilde{\mu}$ at $T=0$, as well as in the critical temperature $T_c$ while remaining below the topological transition. In order to observe universality, we had to use the correlation length as a driving parameter, measure the chemical potential from the bottom of the lowest helicity state, and embody effective-mass renormalization effects of the resonant non-condensed pairs into the critical temperature. Above the topological transition instead, universality is progressively lost. We may argue that the standard BCS-BEC crossover could be replaced by a crossover from a BEC of tightly-bound composite bosons to a BCS of interacting bosonic molecules composed of same-helicity quasi-particles~\cite{Shenoy}. In fact, we can speculate that a calculation of the effective mass or size of the fermion pairs while moving away from the BEC regime, can be used to upgrade the universal parameter $\xi$ with inclusion of a new length which might recover a form of universal behavior for stronger SOC.

Our predictions can be tested in timely experiments with, e.g., effective photon-mediated interactions in optical cavities \cite{Ben_Lev}, and can be interesting in view of applications exploiting topological states as they can be induced by SOC.


\end{document}